# Bound-state solutions of the Klein-Gordon equation for the generalized *PT*-symmetric Hulthén potential

**Harun Egrifes** [1,a] **and Ramazan Sever** [2,b]

[1] Department of Physics, Faculty of Science, Ege University, 35100 Izmir, Turkey

[2] Department of Physics, Faculty of Arts and Sciences, Middle East Technical University, 06531 Ankara, Turkey

## Abstract

The one-dimensional Klein-Gordon equation is solved for the *PT*-symmetric generalized Hulthén potential in the scalar coupling scheme. The relativistic bound-state energy spectrum and the corresponding wave functions are obtained by using the Nikiforov-Uvarov method which is based on solving the second-order linear differential equations by reduction to a generalized equation of hypergeometric type.

*PACS* : 03.65.Fd, 03.65.Ge

*Keywords* : Klein-Gordon equation; Non-Hermitian Hamiltonians; *PT*-symmetry; Pseudo-Hermiticity; Hulthén potential; Bound-states; Exact solutions.

[a] E-mail : harun.egrifes@ege.edu.tr

[b] E-mail : sever@metu.edu.tr

## 1. Introduction

In the last few years there has been considerable work on non-Hermitian Hamiltonians. Among this kind of Hamiltonians, much attention has focused on the investigation of properties of so-called *PT*-symmetric Hamiltonians. Following the early studies of Bender and his co-workers [1], the *PT*-symmetry formulation has been successfully utilized by many authors [2-10]. Non-Hermitian Hamiltonians with real or complex spectra have also been analyzed by using different methods [3-6,10-12]. Non-Hermitian but *PT*-symmetric models have applications in different fields, such as optics [13], nuclear physics [14], condensed matter [15], quantum field theory [16] and population biology [17].

The aim of the present work is to further pursue the development of *PT*-symmetry and to solve the one-dimensional time-independent Klein-Gordon (KG) equation for some complex potentials. In view of the *PT*-symmetric formulation, we will apply the Nikiforov-Uvarov (NU) method [18] to solve the (1+1)-dimensional time-independent KG equation for a spinless particle of rest mass $m$. We have presented exact bound states for a family of exponential-type potentials, i.e., generalized Hulthén potential which can be reduced to the standard Hulthén potential, Woods-Saxon potential and exponential-type screened Coulomb potential. This family of potentials have been successfully applied to a number of different fields of physical systems. Using the quantization of the boundary condition of the states at the origin, Znojil [19] studied another form of the generalized Hulthén potential in non-relativistic and relativistic region. Dominguez-Adame [20] and Chetouani *et al*. [21] also studied relativistic bound states of the standard Hulthén potential. On the other hand, Rao and Kagali and Rao *et al*. [22] investigated relativistic bound states of the exponential-type screened Coulomb potential by means of the one-dimensional KG equation. However, it is well known that for the exponential-type screened Coulomb potential there is no explicit form of the energy expression of bound states for KG [22], Schrödinger [22, 23] and also Dirac equations [24].

In a recent work [25], we have presented the bound-state solutions of the one-dimensional Dirac equation in the vector coupling scheme for *PT*-symmetric potentials with complexified generalized Hulthén potential. In this study we will be dealing with bound-state solutions of the one-dimensional KG equation in the scalar coupling scheme for real and complex forms of generalized Hulthén potential. The organization of the present work is as follows. After a brief introductory discussion of the NU method in Section 2, we obtain the bound-state energy eigenvalues for real and complex cases of generalized Hulthén potential and corresponding eigenfunctions in Section 3. In Section 4, we have presented the NU method for exact bound states of *PT*-symmetric exponential potential. As pointed out by Dutt *et al*. [26], the screened Coulomb potential can very well be represented by an effective Hulthén potential [27]. Finally, conclusions and remarkable facts are discussed in the last section.

## 2. The Nikiforov-Uvarov (NU) method

In recent years, an alternative method known as the NU method has been introduced for solving the non-relativistic and relativistic wave equations. There have been several applications involving the Schrödinger equation with some well-known potentials [28], and the Dirac, KG and Duffin-Kemmer-Petiau equations for the exponential-type potentials using this method [25, 29, 30]. Various special functions appear in the solution of many problems of relativistic and non-relativistic quantum mechanics. The differential equations whose solutions are the special functions can be solved by using the NU method. This method is developed for constructing solutions of the general second-order linear differential equation which are special orthogonal polynomials [18]. It is well known that any given one-dimensional or radial Schrödinger equation and other Schrödinger-like equations can be written as a second-order linear differential equation. Therefore, to apply the NU method one writes it in the generalized hypergeometric differential equation form

$$\psi''(z) + \frac{\tilde{\tau}(z)}{\sigma(z)}\psi'(z) + \frac{\tilde{\sigma}(z)}{\sigma^2(z)}\psi(z) = 0 \tag{1}$$

where $\sigma(z)$ and $\tilde{\sigma}(z)$ are polynomials, at most of second degree, and $\tilde{\tau}(z)$ is a polynomial, at most of first degree.

Using the transformation

$$\psi(z) = \phi(z) y(z) \tag{2}$$

together with the equation determining the eigenvalues

$$\lambda = \lambda_n = -n\tau'(z) - \frac{n(n-1)}{2}\sigma''(z) \qquad (n = 0,1,2,...), \tag{3}$$

Eq.(1) can also be reduced to the following differential equation

$$\sigma(z)y''(z) + \tau(z)y'(z) + \lambda y(z) = 0. \tag{4}$$

This is a differential equation of hypergeometric type, whose polynomial solutions are given by Rodrigues relation [18]

$$y_n(z) = \frac{B_n}{\omega(z)} \frac{d^n}{dz^n}\left[\sigma^n(z)\omega(z)\right] \tag{5}$$

where

$$\frac{\phi'(z)}{\phi(z)} = \frac{\varphi(z)}{\sigma(z)}, \tag{6}$$

$$\tau(z) = \tilde{\tau}(z) + 2\varphi(z), \tag{7}$$

$$\varphi(z) = \frac{\sigma'(z) - \tilde{\tau}(z)}{2} \pm \sqrt{\left(\frac{\sigma'(z) - \tilde{\tau}(z)}{2}\right)^2 - \tilde{\sigma}(z) + k\sigma(z)}, \tag{8}$$

$$k = \lambda - \varphi'(z). \tag{9}$$

Since $\varphi(z)$ has to be a polynomial of degree at most one, in Eq.(8) the expression under the square root must be the square of a polynomial of first degree [18]. This is possible only if its discriminant is zero. Hence we obtain an equation, in the quadratic form, for $k$. After determining $k$, we have $\varphi(z)$ from Eq.(8), and then $\phi(z)$, $\tau(z)$ and $\lambda$, respectively.

**3. Exact bound-state solutions of the generalized Hulthén potential**

It is well known the fact that the exact solutions of the KG equation play an important role in relativistic quantum mechanics. Thus, considerable efforts have been spent in recent years towards obtaining the exact solutions of other types of relativistic wave equations for certain potentials of physical interest [22, 29, 31, 32]. Having given the brief review of the NU method above, let us now consider the (1+1)-dimensional time-independent KG equation for a spinless particle of rest mass $m$

$$\psi''(x) + \frac{1}{\hbar^2 c^2}\left[(E - V(x))^2 - (mc^2 + S(x))^2\right]\psi(x) = 0 \tag{10}$$

in the presence of vector and scalar potentials, where the vector and scalar potentials are given by $V(x)$ and $S(x)$, respectively. We will be dealing with bound-state solutions, i.e., the wave function vanishes at infinity. Here we consider the situation where the scalar potential is greater than the vector potential. Usually it is required that $S(x) > V(x)$ in order to assure the existence of bound states [20, 32, 36].

As is known in relativistic quantum mechanics, the interaction potential can either be introduced in the vector coupling prescription or in the scalar prescription following the minimal coupling rule. While one often deals with the vector coupling prescription, its counterpart, viz., the scalar interaction seldom appears in literature. As a classical application of the scalar prescription, we choose the one-dimensional scalar potential

$$S_q(x) = -S_0 \frac{e^{-\alpha x}}{1 - q e^{-\alpha x}} \tag{11}$$

which is called generalized Hulthén potential [29]. The deformation parameter $q$ determines the shape of the potential. It is worth mentioning here that, for some specific $q$ values this potential transforms to the well-known types : such as for $q = 0$ to the exponential potential, for $q=1$ to the standard Hulthén potential and for $q=-1$ to the Woods-Saxon potential. When $\alpha \to 0$, the potential is close to the origin

$$S_q(x) \approx \frac{S_0}{q-1} + \frac{S_0}{(q-1)^2} \alpha x \qquad (12)$$

and behaves like a linear potential [33] which is a Lorentz scalar used in the study of quarkonium systems, where $\alpha$ denotes the range parameter and $S_0$ denotes the scalar coupling constant. Such a potential arises, for example, for charged bosons in the electric field of two paralel condenser plates separated by a distance $\ell$ [34]. It also arises in an approximate treatment of particle production in QCD. When a pair of quarks and an antiquark $q\bar{q}$ are stretched, the field between the quark and the antiquark is represented as an approximation by an Abelian gauge field. The strength parameter $\frac{S_0}{(q-1)^2}\alpha$ is then related to the string tension [35]. The mixed vector-scalar interaction has also been analyzed in one-plus-one dimensions for a linear potential [36].

For $V(x)=0$, the one-dimensional KG equation for a given general potential $S(x)$ in the scalar coupling scheme reads

$$\psi''(x) + \frac{1}{\hbar^2 c^2}\left[E^2 - \left(mc^2 + S(x)\right)^2\right]\psi(x) = 0. \qquad (13)$$

It is worth to note that, in this case ( i.e., the case of a pure scalar potential ) one finds energy levels for particles and antiparticles arranged symmetrically about $E=0$ [37]. Eq.(13) can readily be transformed to the Schrödinger-like second-order differential equation,

$$\psi''(x) + \frac{2m}{\hbar^2}\left[E_{\text{eff}} - U_{\text{eff}}(x)\right]\psi(x) = 0 \qquad (14)$$

with an "effective energy" $E_{\text{eff}}$ and "effective potential" $U_{\text{eff}}(x)$ given by

$$E_{\text{eff}} = \frac{E^2 - m^2 c^4}{2mc^2}, \quad U_{\text{eff}}(x) = \frac{S^2(x)}{2mc^2} + S(x). \qquad (15)$$

It is straightforward to see that Eq.(13) can be written as

$$\psi''(x) + \left[-S^2(x) - 2mS(x) - (m^2 - E^2)\right]\psi(x) = 0, \tag{16}$$

where we have used the natural units ($\hbar = c = 1$).

By substituting Eq.(11) into Eq.(16) and defining a new variable $z = S_0 e^{-\alpha x}$, one obtains the generalized equation of hypergeometric type which is given by Eq.(1)

$$\psi_q''(z) + \frac{S_0 - qz}{z(S_0 - qz)}\psi_q'(z) + \frac{1}{[z(S_0 - qz)]^2}\left[-\left(\gamma^2 + q\beta^2 + q^2\varepsilon^2\right)z^2 + S_0\left(\beta^2 + 2q\varepsilon^2\right)z - S_0^2\varepsilon^2\right]\psi_q(z) = 0 \tag{17}$$

for which

$$\tilde{\tau}_q(z) = S_0 - qz, \sigma_q(z) = z(S_0 - qz), \tilde{\sigma}_q(z) = -\left(\gamma^2 + q\beta^2 + q^2\varepsilon^2\right)z^2 + S_0\left(\beta^2 + 2q\varepsilon^2\right)z - S_0^2\varepsilon^2,$$

$$\gamma^2 = \frac{S_0^2}{\alpha^2}, \quad \beta^2 = \frac{2mS_0}{\alpha^2}, \quad \varepsilon^2 = \frac{1}{\alpha^2}\left(m^2 - E^2\right), \tag{18}$$

with real $\varepsilon^2 \geq 0$ ($E^2 \leq m^2$) for bound states [20]. Substituting $\sigma_q(z), \tilde{\tau}_q(z)$ and $\tilde{\sigma}_q(z)$ into Eq.(8), one finds

$$\varphi_q(z) = -\frac{qz}{2} \pm \frac{1}{2}\sqrt{\left[q^2 + 4\left(\gamma^2 + q\beta^2 + q^2\varepsilon^2\right) - 4qk\right]z^2 + 4S_0\left[k - \left(\beta^2 + 2q\varepsilon^2\right)\right]z + 4S_0^2\varepsilon^2}. \tag{19}$$

The constant parameter $k$ can be determined from the condition that the expression under the square root has a double zero, i.e., $k_\pm = \beta^2 \pm \sqrt{q^2 + 4\gamma^2}\,\varepsilon$. We then obtain the following possible forms of $\varphi_q(z)$:

$$\varphi_q(z) = -\frac{qz}{2} \pm \begin{cases} \frac{1}{2}\left[(a - 2q\varepsilon)z + 2S_0\varepsilon\right] & \text{for } k_+ = \beta^2 + a\varepsilon \\ \frac{1}{2}\left[(a + 2q\varepsilon)z - 2S_0\varepsilon\right] & \text{for } k_- = \beta^2 - a\varepsilon \end{cases} \tag{20}$$

where $a = \sqrt{q^2 + 4\gamma^2} = \sqrt{q^2 + 4(S_0^2/\alpha^2)}$. The polynomial $\varphi_q(z)$ is chosen such that the function $\tau_q(z)$ given by Eq.(7) will have a negative derivative [18]. This condition is satisfied by

$$\tau_q(z) = (1 + 2\varepsilon)S_0 - (2q + a + 2q\varepsilon)z, \quad \tau'_q = -(2q + a + 2q\varepsilon) < 0 \qquad (21)$$

which corresponds to

$$\varphi_q(z) = S_0\varepsilon - \frac{1}{2}(q + a + 2q\varepsilon)z \qquad (22)$$

for $k_- = \beta^2 - a\varepsilon$. Then, we have another constant, $\lambda = k_- + \varphi'_q(z)$, written as

$$\lambda = \beta^2 - \frac{1}{2}(a + q) - (a + q)\varepsilon. \qquad (23)$$

Thus, substituting $\lambda$, $\tau'_q(z)$ and $\sigma''_q(z)$ into Eq.(3), the exact energy eigenvalues of the generalized Hulthén potential are determined as

$$E_n(q, \alpha, S_0) = \pm \frac{1}{4q\kappa_n(q, \alpha, S_0)} \sqrt{\left(\kappa_n^2(q, \alpha, S_0) - 4S_0^2\right)\left((2S_0 + 4qm)^2 - \kappa_n^2(q, \alpha, S_0)\right)} \qquad (24)$$

with $\kappa_n(q, \alpha, S_0) = \sqrt{q^2\alpha^2 + 4S_0^2} + q\alpha(2n + 1)$. For pure attractive scalar potential, all bound states appear in pairs, with energies $\pm E_n$. Since the KG equation is independent of the sign of $E$ for scalar potentials, the wave functions become the same for both energy values. It should be noted that the result given in Eq.(24) is consistent with that given in Eq.(20) obtained by means of supersymmetric method in Ref. [38].

To find the function $y(z)$, which is the polynomial solution of hypergeometric-type equation, we multiply Eq.(4) by an appropriate function $\omega(z)$ so that it can be written in self-adjoint form [18]

$$(\sigma \omega y')' + \lambda \omega y = 0. \qquad (25)$$

Here $\omega(z)$ satisfies the differential equation $(\sigma \omega)' = \tau \omega$ which yields

$$\omega_q(z) = z^{2\varepsilon} (S_0 - qz)^{a/q}. \tag{26}$$

We thus obtain the eigenfunctions of hypergeometric-type equation from the Rodrigues relation given by Eq.(5) in the following form :

$$y_{nq}(z) = B_{nq}\, z^{-2\varepsilon} (S_0 - qz)^{-a/q} \frac{d^n}{dz^n}\left[ z^{n+2\varepsilon} (S_0 - qz)^{n+(a/q)} \right]. \tag{27}$$

The eigenfunctions $y_{nq}(z)$ are, up to a numerical factor, the Jacobi polynomials $P_n^{(2\varepsilon, a/q)}(s)$ with $s = 1 - \frac{2q}{S_0} z$ [18, 39]. By substituting $\varphi_q(z)$ and $\sigma_q(z)$ in Eq.(6), one can find the other factor of the wave function giving

$$\phi_q(z) = z^{\varepsilon} (S_0 - qz)^{(a+q)/2q}. \tag{28}$$

As stated in Eq.(2), the wave function is constructed as a product of two independent parts. In this case the wave functions $\psi_{nq}(z)$ can be determined as

$$\psi_{nq}(z) = \phi_q(z)\, y_{nq}(z)$$

$$= C_{nq}\, z^{\varepsilon} (S_0 - qz)^{(a+q)/2q}\, P_n^{(2\varepsilon, a/q)}\left(1 - \frac{2q}{S_0} z\right) \tag{29}$$

where $z = S_0\, e^{-\alpha x}$ and $C_{nq}$ being a normalization constant. Notice that $\psi_{nq}(x)$ decreases exponentially as $x \to \infty$, being square-integrable and thus representing a truly bound state.

### 3.1 *Non-Hermitian PT-symmetric generalized Hulthén potential*

The Hermiticity of a Hamiltonian was supposed to be necessary condition for the real spectrum until the year 1998 [1]. A conjecture due to Bender and Boetcher has relaxed this condition by introducing the concept of *PT*-symmetric Hamiltonians. A Hamiltonian is said to be *PT*-symmetric when $[H, PT] = 0$, where $P$ denotes parity operator (space reflection), i.e., $P: f(x) \to f(-x)$, and $T$ denotes time-reversal, i.e., $T: i \to -i$, namely, for a given potential $S(x)$, when one makes the transformation of $x \to -x$ (or $x \to \xi - x$) and $i \to -i$, if the relation $S(-x) = S^*(x)$ or $S(\xi - x) = S^*(x)$ exists, then the potential $S(x)$ is said to be *PT*-symmetric.

Now let us consider the case, namely, at least one of the potential parameters is complex. If $\alpha$ is a pure imaginary parameter, i.e. $\alpha \to i\alpha$, such potentials are written as a complex function

$$S_q(x) = \frac{S_0}{q^2 - 2q\cos(\alpha x) + 1}\left[q - \cos(\alpha x) + i\sin(\alpha x)\right] \qquad (30)$$

which is *PT*-symmetric but non-Hermitian. It is worthwhile here to point out that, such as a complex periodic potential having *PT*-symmetry of the form $V(x) = i\sin^{2n+1}(x)$ $(n = 0,1,2,...)$ which exhibit real band spectra was discussed in detail by Bender *et al*. [40]. The complex potential (30) has real spectra given by

$$E_n(q,\alpha,S_0) = \pm \frac{1}{4q\mu_n(q,\alpha,S_0)}\sqrt{\left(\mu_n^2(q,\alpha,S_0) + 4S_0^2\right)\left((2S_0 + 4qm)^2 + \mu_n^2(q,\alpha,S_0)\right)} \qquad (31)$$

where $\mu_n(q,\alpha,S_0) = \sqrt{q^2\alpha^2 - 4S_0^2} + q\alpha(2n+1)$. If $q^2\alpha^2 \geq 4S_0^2$, there exist bound state, otherwise there are no bound states.

Referring back to Eq.(29), the corresponding wave functions $\psi_{nq}(z)$ are identified in the form

$$\psi_{nq}(z) = C_{nq}\, z^{-i\varepsilon}\, (S_0 - qz)^{(b+q)/2q}\, P_n^{(2i\varepsilon,\, b/q)}\!\left(1 - \frac{2q}{S_0}z\right) \qquad (32)$$

with $b = \sqrt{q^2 - 4\gamma^2}$ and $z = S_0\, e^{-i\alpha x}$.

Fig.1 illustrates the ground-state energy level as a function of the coupling constant $S_0$ for various shape parameters. The range parameter $\alpha$ is chosen to be $\alpha = 1/\lambda_c$, where $\lambda_c = \hbar/mc = 1/m$ ($\hbar = 1, c = 1$) denotes the Compton wavelength of the KG particle. It can be seen easily that, while $S_0 \to 0$ in the ground-state (i.e., $n = 0$), all energy eigenvalues tend to the value $E_0 \approx 1.118\, m$. As it can be seen from Table 1, *PT*-symmetric non-Hermitian generalized Hulthén potential generates real and positive bound states. For fixed $S_0 = 0.25\, m$ and any given $\alpha$, all the binding energies $E_b = E_n - mc^2$ are decreasing with increasing $q$. The values obtained in Ref.[29] using the vector potential are also shown in Table 1 for comparison. Fig.2 shows the variation of the energy eigenvalues of a KG particle, which is moving in *PT*-symmetric potential given by Eq.(30), as a function of the range parameter $\alpha$ for $q = 1.0$. The coupling strength parameter $S_0$ is chosen to be $S_0 = 0.5\, m$.

### *3.2 Pseudo-Hermiticity and PT-symmetry*

It is interesting to note that when all three parameters $V_0$, $q$ and $\alpha$ are pure imaginary at the same time, i.e., $S_0 \to i S_0$, $q \to i q$ and $\alpha \to i \alpha$, the potential given by Eq.(11) transforms to the form

$$S_q(x) = \frac{S_0}{q^2 - 2q\sin(\alpha x) + 1}\left[q - \sin(\alpha x) - i\cos(\alpha x)\right]. \tag{33}$$

This form of the potential has a $\pi/2$ phase difference with respect to the potential given by Eq.(30). Recently, Mostafazadeh [8] has shown that the potentials of this form are *P*-pseudo-Hermitian and claimed that the $\eta$-pseudo-Hermiticity, $\eta H \eta^{-1} = H^+$, is the necessary condition for having real spectrum, where $\eta$ is referred to as a Hermitian linear automorphism. For a non-Hermitian potential $S(x)$, the necessary and sufficient condition for having a real energy spectrum is that there exist an invertible linear operator $O$ such that $S(x)$ is $\eta$-pseudo-Hermitian, where $\eta = O O^+$. A potential $S(x)$ is said to hold $\eta$-pseudo-Hermiticity when $\eta S(x) \eta^{-1} = S^*(x)$. Ahmed [41] has suggested an explicit form for a Hermitian linear automorphism, $\eta = e^{-\theta p}$, $p = -i\, d/dx$, which affects an imaginary shift of the coordinate: $\eta x \eta^{-1} = x + i\theta$.

If we replace $x$ by $\left(\dfrac{\pi}{2\alpha} - x\right)$, we have $\sin(\alpha x) \to \cos(\alpha x)$ and also $\cos(\alpha x) \to \sin(\alpha x)$. Thus we obtain $P\, S_q(x)\, P^{-1} = S_q^*(x)$ for the complex potential given by Eq.(33) where the Hermitian, linear and invertible operator $\eta$ is the the parity operator $P$, which acts on the position operator as $P\, x\, P^{-1} = \dfrac{\pi}{2\alpha} - x$ [42]. Hence, the complex version of the generalized Hulthén potential possesses $P$-pseudo-Hermiticity. Under joint action of spatial reflection $\left(P: x \to \dfrac{\pi}{2\alpha} - x\right)$ and time-reversal $(T: i \to -i)$, we obtain $PT\, S_q(x)\,(PT)^{-1} = S_q(x)$ for the complex version of the potential function given by Eq.(11). Therefore, we can say that the complex potential given by Eq.(33) also holds $PT$-symmetry. It has exact real spectra

$$E_n(q,\alpha,S_0) = \pm \frac{1}{4q\delta_n(q,\alpha,S_0)} \sqrt{\left(\delta_n^2(q,\alpha,S_0) + 4S_0^2\right)\left((2S_0 + 4qm)^2 + \delta_n^2(q,\alpha,S_0)\right)} \quad (34)$$

where $\delta_n(q,\alpha,S_0) = \sqrt{q^2\alpha^2 - 4S_0^2} - q\alpha(2n+1)$ with $q^2\alpha^2 \geq 4S_0^2$.

Now again referring back to Eq.(29), the corresponding wave functions $\psi_{nq}(z)$ are identified in the form

$$\psi_{nq}(z) = C_{nq}\, z^{-i\varepsilon}\, (i S_0 - i q z)^{(b+q)/2q}\, P_n^{(2i\varepsilon,\, b/q)}\!\left(1 - \frac{2q}{S_0} z\right) \quad (35)$$

where $z = i S_0\, e^{-i\alpha x}$.

### 4. *PT*-symmetric exponential potential

In the previous section we have obtained the bound-state solutions of the generalized Hulthén potential with $q \neq 0$ and presented the explicit form of the eigenvalues and the wave functions. Now we turn our attention to the $q = 0$ case. Note that, for $q = 0$, there is no explicit form of the energy expression of bound states for KG [22], Schrödinger [22, 23] and also Dirac equations [24].

Inserting the *PT*-symmetric exponential potential $S(x) = -S_0 e^{-i\alpha x}$ in Eq.(16), we obtain

$$\psi''(x) + \left[ -S_0^2 e^{-i2\alpha x} + 2mS_0 e^{-i\alpha x} - (m^2 - E^2) \right] \psi(x) = 0. \tag{36}$$

Defining a new variable $s = S_0 e^{-i\alpha x}$, Eq.(36) transforms to

$$\psi''(s) + \frac{1}{s}\psi'(s) + \frac{1}{s^2}\left[ \frac{1}{\alpha^2} s^2 - \frac{2m}{\alpha^2} s + \varepsilon^2 \right] \psi(s) = 0 \tag{37}$$

and the corresponding $\varphi(s)$ is determined as

$$\varphi(s) = \pm \begin{cases} \dfrac{i}{\alpha}(s - \alpha\varepsilon) & \text{for} \quad k_+ = -\dfrac{2m}{\alpha^2} + \dfrac{2\varepsilon}{\alpha}, \\[2ex] \dfrac{i}{\alpha}(s + \alpha\varepsilon) & \text{for} \quad k_- = -\dfrac{2m}{\alpha^2} - \dfrac{2\varepsilon}{\alpha}. \end{cases} \tag{38}$$

Following a procedure similar to the previous case, when $\varphi(s) = -\dfrac{i}{\alpha} s - i\varepsilon$ is chosen for

$$k_- = -\frac{2m}{\alpha^2} - \frac{2\varepsilon}{\alpha},$$

$$\tau(s) = (1 - 2i\varepsilon) - i\frac{2}{\alpha}s \ , \ \lambda = -\frac{2m}{\alpha^2} - \frac{2\varepsilon}{\alpha} - \frac{i}{\alpha} \ , \ \phi(s) = s^{-i\varepsilon} e^{-is/\alpha} \tag{39}$$

is obtained. Substituting $\sigma(s)$ and $\tau(s)$, together with $\lambda$, into Eq.(4) yields

$$s\, y''(s) + \left[ (1 - 2i\varepsilon) - \frac{2i}{\alpha}s \right] y'(s) - \left( \frac{i}{\alpha} + \frac{2\varepsilon}{\alpha} + \frac{2m}{\alpha^2} \right) y(s) = 0. \tag{40}$$

Further by putting $y(z) = e^{z/2} z^{-\nu-(1/2)} u(z)$, Eq.(40) can also be reduced to the standard Whittaker differential equation [43] :

$$u''(z) + \left[ -\frac{1}{4} + \frac{\mu}{z} + \frac{\frac{1}{4} - \nu^2}{z^2} \right] u(z) = 0 \tag{41}$$

whose solutions vanishing at infinity can be written in terms of the Kummer confluent hypergeometric functions,

$$u(z) = M_{\mu,\nu}(z) = e^{-z/2} z^{\nu+(1/2)} {}_1F_1\left(\frac{1}{2} + \nu - \mu; 1 + 2\nu; z\right) \quad (42)$$

where $\mu = im/\alpha$, $\nu = -i\varepsilon$, and $z = \frac{2i}{\alpha} s = \frac{2i}{\alpha} S_0 e^{-i\alpha x}$.

Two linearly independent series solutions to Eq.(40) are given by in terms of confluent hypergeometric functions as follows :

$$y(z) = A \, {}_1F_1\left(\frac{1}{2} - i\varepsilon - i\frac{m}{\alpha}; 1 - 2i\varepsilon; z\right) + B \, z^{2i\varepsilon} \, {}_1F_1\left(\frac{1}{2} + i\varepsilon - i\frac{m}{\alpha}; 1 + 2i\varepsilon; z\right). \quad (43)$$

In this solution the second term is physically unacceptable since

$$z^{2i\varepsilon} = \left(\frac{2i}{\alpha} S_0 e^{-i\alpha x}\right)^{2i\varepsilon} = \left(\frac{2i}{\alpha} S_0\right)^{2i\varepsilon} e^{2\alpha\varepsilon x} \to \infty \quad \text{when} \quad x \to \infty. \text{ Thus, the acceptable}$$

solution vanishing at infinity is found to be

$$\psi(s) = \phi(s) y(s) = A \, s^{-i\varepsilon} e^{-is/\alpha} \, {}_1F_1\left(\frac{1}{2} - i\varepsilon - i\frac{m}{\alpha}; 1 - 2i\varepsilon; \frac{2i}{\alpha} s\right) \quad (44)$$

A being a normalization constant. In terms of the original variable $x$, we can write the wave function as

$$\psi(x) = A \, S_0^{-i\varepsilon} \exp\left(-i\frac{S_0}{\alpha} e^{-i\alpha x} - \alpha\varepsilon x\right) {}_1F_1\left(\frac{1}{2} - i\varepsilon - i\frac{m}{\alpha}; 1 - 2i\varepsilon; \frac{2i}{\alpha} S_0 e^{-i\alpha x}\right). \quad (45)$$

To obtain the energy eigenvalues we demand that $\psi(x)$ vanishes at the origin ($x = 0$). This implies the vanishing of the Kummer function at the origin, thus leading to the following eigenvalue equation

$$_1F_1\left(\frac{1}{2} - i\varepsilon - i\frac{m}{\alpha}; 1 - 2i\varepsilon; \frac{2i}{\alpha} S_0\right) = 0 \quad (46)$$

which is an implicit equation for the determination of the energy eigenvalues. The explicit solution of Eq.(46), showing the dependence of the energy eigenvalues $\varepsilon = \frac{\sqrt{m^2 - E^2}}{\alpha}$ on $S_0$ and $\alpha$, is then reduced to find the zeros of the Kummer confluent hypergeometric function. These are complicated transcendental equations, the real roots of which can only be found numerically, using computational tools like Mathematica or Maple [22, 44].

## 5. Conclusions

We have found that the (1+1)-dimensional time independent Klein-Gordon equation for the generalized scalar Hulthén potential can be solved exactly. The relativistic bound-state energy spectrum and the corresponding wave functions have been obtained by the NU method when the scalar coupling is of sufficient intensity compared to the vector coupling. While the relativistic bound-state eigenfunctions are expressed in terms of Jacobi polynomials for $q \neq 0$, they are expressed in terms of confluent hypergeometric functions for $q = 0$. Some interesting results including complex *PT*-symmetric and pseudo-Hermitian versions of the generalized Hulthén potential have also been discussed. We show that it is possible to obtain relativistic bound states of complex quantum mechanical formulation.

## References

[1] C.M. Bender, S. Boettcher, Phys.Rev.Lett. 80 (1998) 5243;

   C.M. Bender, S. Boettcher, P.N. Meisenger, J.Math.Phys. 40 (1999) 2201.

[2] C.M. Bender, G.V. Dunne, J.Math.Phys. 40 (1999) 4616;

   G.A. Mezincescu, J.Phys. A : Math. Gen. 33 (2000) 4911;

   E. Delabaere, D.T. Trinh, J.Phys. A : Math. Gen. 33 (2000) 8771;

   M. Znojil, M. Tater, J.Phys. A : Math. Gen. 34 (2001) 1793;

   C.M. Bender, G.V. Dunne, P.N. Meisenger, M. Simsek, Phys.Lett. A 281 (2001) 311;

   Z. Ahmed, Phys.Lett. A 282 (2001) 343;

   Z. Ahmed, Phys.Lett. A 284 (2001) 231;

   P. Dorey, C. Dunning, R. Tateo, J.Phys. A : Math. Gen. 34 (2001) 5679;

   B. Bagchi, C. Quesne, Mod.Phys.Lett. A 16 (2001) 2449;

   K.C. Shin, J.Math.Phys. 42 (2001) 2513;

   G.S. Japaridze, J.Phys. A : Math. Gen. 35 (2002) 1709;

   C.S. Jia, S.C. Li, Y.Li, L.T. Sun, Phys.Lett. A 300 (2002) 115;

   C.S. Jia, L.Z. Yi, Y. Sun, J.Y. Liu, L.T. Sun, Mod.Phys.Lett. A 18 (2003) 1247;

   C.S. Jia, Y. Li, Y. Sun, J.Y. Liu, L.T. Sun, Phys.Lett. A 311 (2003) 115;

   L.Z. Yi, Y.F. Diao, J.Y. Liu, C.S. Jia, Phys.Lett. A 333 (2004) 212;

   C.S. Jia, L.Z. Yi, Y.Q. Zhao, J.Y. Liu, L.T. Sun, Mod.Phys.Lett. A 20 (2005) 1753.

[3] F. Cannata, G. Junker, J. Trost, Phys.Lett. A 246 (1998) 219.

[4] A. Khare, B.P. Mandal, Phys.Lett. A 272 (2000) 53.

[5] B.Bagchi, C. Quesne, Phys.Lett. A 273 (2000) 285.

[6] Z. Ahmed, Phys.Lett. A 282 (2001) 343.

[7] L. Solombrino, J.Math.Phys. 43 (2002) 5439.

[8] A. Mostafazadeh, J.Math.Phys. 43 (2002) 205;

A. Mostafazadeh, J.Math.Phys. 43 (2002) 2814;

A. Mostafazadeh, J.Math.Phys. 43 (2002) 3944.

[9] M.Znojil, Phys.Lett. A 264 (1999) 108.

[10] B. Bagchi, C. Quesne, Phys.Lett. A 300 (2002) 18.

[11] C.M. Bender, M. Berry, P.N. Meisenger, V.M. Savage, M. Simsek, J.Phys. A : Math. Gen. 34 (2001) L31.

[12] C.M. Bender, E.J Weniger, J.Math.Phys. 42 (2001) 2167;

C.M. Bender, S. Boettcher, H.F. Jones, P.N. Meisenger, M. Simsek, Phys.Lett. A 291 (2001) 197.

[13] S.K. Moayedi, A. Rostami, Eur.Phys.J. B 36 (2003) 359.

[14] D. Baye, G. Levai, J.M. Sparenberg, Nucl.Phys. A 599 (1996) 435;

R.N. Deb, A. Khare, B.D. Roy, Phys.Lett. A 307 (2003) 215.

[15] N. Hatano, D.R. Nelson, Phys.Rev.Lett. 77 (1996) 570;

N. Hatano, D.R. Nelson, Phys.Rev.B 56 (1997) 8651.

[16] C.M. Bender, K.A. Milton, V.M. Savage, Phys.Rev. D 62 (2000) 085001;

C. Bernard, V.M. Savage, Phys.Rev. D 64 (2001) 085010.

[17] D.R.Nelson, N.M. Shnerb, Phys.Rev.E 58 (1998) 1383.

[18] A.F. Nikiforov, V.B. Uvarov, Special Functions of Mathematical Physics, Birkhauser, Basel, 1988.

[19] M. Znojil, J.Phys. A : Math. Gen. 14 (1981) 383.

[20] F. Dominguez-Adame, Phys.Lett. A 136 (1989) 175.

[21] L. Chetouani, L Guechi, A. Lecheheb, T.F. Hammann, A. Messouber, Physica A 234 (1996) 529.

[22] N.A. Rao, B.A. Kagali, Phys.Lett. A 296 (2002) 192;

N.A. Rao, B.A. Kagali, V. Sivramkrishna, Int.J.Mod.Phys. A 17 (2002) 4793.

[23] S. Flügge, Practical Quantum Mechanics, Springer, Berlin, 1974.

[24] F. Dominguez-Adame, A. Rodriguez, Phys.Lett. A 198 (1995) 275;

B.A. Kagali, N.A. Rao, V. Sivramkrishna, Mod.Phys.Lett. A 17 (2002) 2049;

V.M. Villalba, W.Greiner, Phys.Rev.A 67 (2003) 052707;

A.S. de Castro, M. Hott, Phys.Lett. A 342 (2005) 53.

[25] H. Egrifes, R. Sever, Phys.Lett. A 344 (2005) 117.

[26] R. Dutt, K. Chowdhury, Y.P Varshni, J.Phys. A : Math. Gen. 18 (1985) 1379.

[27] R.L. Greene, C. Aldrich, Phys.Rev. A 14 (1976) 2363.

[28] M. Aktas, R. Sever, J.Math.Chem. 37 (2005) 139;

C. Berkdemir, A. Berkdemir, R. Sever, Phys.Rev. C 72 (2005) 027001;

F. Yasuk, C. Berkdemir, A. Berkdemir, J.Phys. A : Math. Gen. 38 (2005) 6579.

[29] M. Simsek, H. Egrifes, J.Phys. A : Math. Gen. 37 (2004) 4379.

[30] F. Yasuk, C. Berkdemir, A. Berkdemir, C. Onem, Physica Scripta 71 (2005) 340.

[31] C.F. Hou, Z.X. Zhou, Y. Li, Acta Physica Sinica 8 (1999) 561, overseas edition;

W.C. Qiang, Chinese Phys. 11 (2002) 757;

W.C. Qiang, Chinese Phys. 12 (2003) 1054;

S.H. Dong, X.Y. Gu, Z.Q. Ma, J. Yu, Int.J.Mod.Phys. E 12 (2003) 555;

O.Mustafa, J.Phys. A : Math.Gen. 36 (2003) 5067;

W.C. Qiang, Chinese Phys. 13 (2004) 571;

W.C. Qiang, Chinese Phys. 13 (2004) 575;

Z.Q. Ma, S.H. Dong, X.Y. Gu, J. Yu, M. Lozada-Cassou, Int.J.Mod.Phys. E 13 (2004) 597;

L.Z.Yi, Y.F. Diao, J.Y. Liu, C.S. Jia, Phys.Lett. A 333 (2004) 212;

Y.F. Diao, L.Z. Yi, C.S. Jia, Phys.Lett. A 332 (2004) 157;

G. Chen, Z.D. Chen, Z.M. Lou, Chinese Phys. 13 (2004) 279;

X.A. Zheng, K. Chen, Z.L. Duan, Chinese Phys. 14 (2005) 42;

A.S. de Castro, Phys.Lett. A 338 (2005) 81;

A.S. de Castro, Phys.Lett. A 346 (2005) 71;

X.Q. Zhao, C.S. Jia, Q.B. Yang, Phys.Lett. A 337 (2005) 189;

C. Rojas, V.M. Villalba, Phys.Rev. A 71 (2005) 052101;

A.D. Alhaidari, H. Bahlouli, A. Al-Hasan, Phys.Lett. A 349 (2006) 87;

G. Chen, Z.D. Chen, P.C. Xuan, Phys.Lett. A 352 (2006) 317.

[32] G. Chen, Mod.Phys.Lett. A 19 (2004) 2009;

G. Chen, Phys.Lett. A 339 (2005) 300;

A.S. de Castro, Phys.Lett. A 338 (2005) 81;

A. de Souza Dutra, G. Chen, Phys.Lett. A 349 (2006) 297.

[33] B.Ram, Am.J.Phys. 50 (1982) 549;

R.K. Su, Y. Zhang, J.Phys.A : Math. Gen. 17 (1984) 851;

R.K. Roychoudhury, Y.P. Varshni, J.Phys. A : Math. Gen. 20 (1987) L1083;

O. Mustafa, R. Sever, Phys.Rev. A 44 (1991) 4142;

R.S. Bhalerao, B. Ram, Am.J.Phys. 69 (2001) 817;

J.R. Hiller, Am.J.Phys. 70 (2002) 522;

A.S. de Castro, Phys.Lett. A 305 (2002) 100.

[34] R.C. Wang, C.Y. Wong, Phys.Rev. D 38 (1988) 348.

[35] A. Casher, H. Neuberger, S. Nussinov, Phys.Rev. D 20 (1979) 179.

[36] A.S. de Castro, Phys.Lett. A 305 (2002) 100.

[37] F.A.B. Coutinho, Y. Nogami, Phys.Lett. A 124 (1987) 211;

F.A.B. Coutinho, Y. Nogami, F.M. Toyama, Am.J.Phys. 56 (1988) 904.

[38] G. Chen, Z.D. Chen, Z.M. Lou, Phys.Lett. A 331 (2004) 374.

[39] W. Magnus, F. Oberhettinger, R.P. Soni, Formulas and Theorems for the Special Functions of Mathematical Physics, third ed., Springer, Berlin, 1966.

[40] C.M. Bender, G.V. Dunne, P.N. Meisinger, Phys.Lett.A 252 (1999) 272.

[41] Z. Ahmed, Phys.Lett. A 290 ( 2001) 19.

[42] C.S. Jia, P.Y. Lin, L.T. Sun, Phys.Lett. A 298 ( 2002) 78;

C.S. Jia, Y. Sun, Y.Li, Phys.Lett. A 305 (2002) 231.

[43] M. Abromowitz, I. Stegun, Handbook of Mathematical Functions with Formulas, Graphs and Mathematical Tables, Dover, New York, 1964.

[44] W. Greiner, Relativistic Quantum Mechanics, Springer-Verlag, Berlin, 1990.

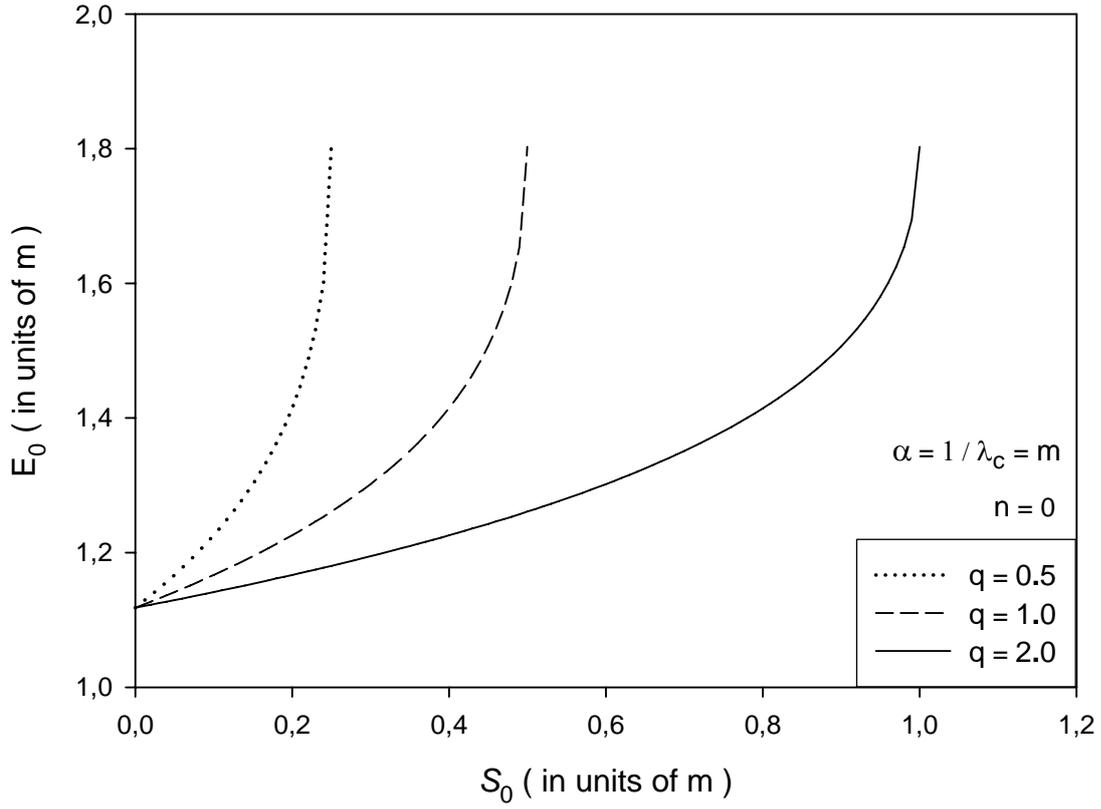

**Fig.1** The variation of the ground-state ($n = 0$) energy of a Klein-Gordon particle, which is moving in *PT*-symmetric potential given by Eq.(30), as a function of the coupling constant $S_0$ for three different shape parameters. The constant $\alpha$ characterizing the range of the potential is $\alpha = 1/\lambda_c = 1/\hbar/mc = m$ ($\hbar = 1, c = 1$).

Figure 2

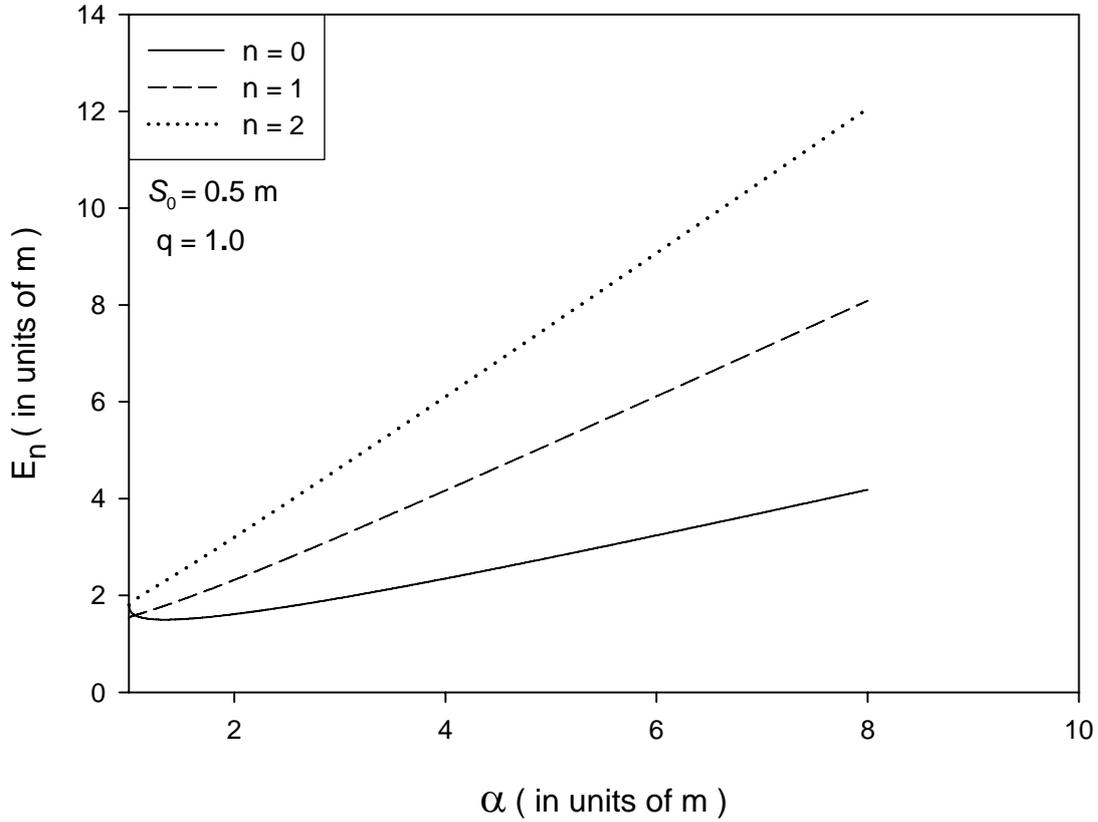

**Fig.2** The variation of the energy eigenvalues of a Klein-Gordon particle, which is moving in *PT*-symmetric potential given by Eq.(30), as a function of the range parameter $\alpha$ for $q = 1.0$. The coupling strength $S_0$ of the potential is chosen to be $S_0 = 0.5\,m$. The curves are plotted for the first three values of the vibrational quantum number $n$.

**Table 1.** The ground-state binding energies ($E_b = E_0 - mc^2$) of a KG particle of mass unity as a function of $q$ for various values of the range parameter $\alpha$ in *PT*-symmetric potential given by Eq.(30) ($S_0 = 0.25$). The values obtained in Ref.[29] using the vector potential are also shown in table. Here $\alpha$ is expressed in units of the Compton wavelength ($\alpha = 1/\lambda_c = mc/\hbar$).

| | Vector potential (Ref.[29]) | | | Scalar potential | | |
|---|---|---|---|---|---|---|
| | $\alpha = 0.5$ | $\alpha = 1.0$ | $\alpha = 2.0$ | $\alpha = 0.5$ | $\alpha = 1.0$ | $\alpha = 2.0$ |
| $q$ | $E_0 - mc^2$ | $E_0 - mc^2$ | $E_0 - mc^2$ | $E_0 - mc^2$ | $E_0 - mc^2$ | $E_0 - mc^2$ |
| 0.5 | 0.584789 | 0.503548 | 0.726663 | ----------- | 0.802776 | 0.614831 |
| 1.0 | 0.264375 | 0.282323 | 0.555525 | 0.600781 | 0.260846 | 0.506699 |
| 1.5 | 0.168081 | 0.219446 | 0.504860 | 0.182955 | 0.204579 | 0.474727 |
| 2.0 | 0.125097 | 0.190836 | 0.480840 | 0.126170 | 0.180312 | 0.459301 |